\documentclass{article}

\usepackage{spconf,amsmath,graphicx,hyperref,xcolor,amsfonts}

\hypersetup{
    colorlinks=true,
    citecolor=blue,
    linkcolor=blue,
    filecolor=magenta,      
    urlcolor=blue,
}
\usepackage{enumitem}
\setlist[1]{itemsep=1pt}

\usepackage{booktabs}
\graphicspath{{figures/}} 

\title{ICASSP 2022 Acoustic Echo Cancellation Challenge}
\name{
\parbox{\linewidth}{\centering
Ross Cutler, Ando Saabas, Tanel Parnamaa, Marju Purin, Hannes Gamper, Sebastian Braun, Karsten Sørensen, Robert Aichner}}
\address{Microsoft Corporation, Redmond, USA \\
    firstname.lastname@microsoft.com}
\begin{document}
\ninept
\maketitle
\begin{abstract}
The ICASSP 2022 Acoustic Echo Cancellation Challenge is intended to stimulate research in acoustic echo cancellation (AEC), which is an important area of speech enhancement and still a top issue in audio communication. This is the third AEC challenge and it is enhanced by including mobile scenarios, adding speech recognition word accuracy rate as a metric, and making the audio 48 kHz. We open source two large datasets to train AEC models under both single talk and double talk scenarios. These datasets consist of recordings from more than 10,000 real audio devices and human speakers in real environments, as well as a synthetic dataset. We open source an online subjective test framework and provide an online objective metric service for researchers to quickly test their results. The winners of this challenge were selected based on the average Mean Opinion Score (MOS) achieved across all scenarios and the word accuracy rate.
\end{abstract}
\vspace{2mm}

\section{Introduction}
\label{sec:intro}
With the growing popularity and need for working remotely, the use of teleconferencing systems such as Microsoft Teams, Skype, WebEx, Zoom, etc., has increased significantly. It is imperative to have good quality calls to make the user's experience pleasant and productive. The degradation of call quality due to acoustic echoes is one of the major sources of poor speech quality ratings in voice and video calls. While digital signal processing (DSP) based AEC models have been used to remove these echoes during calls, their performance can degrade when model assumptions are violated, e.g., fast time-varying acoustic conditions, unknown signal processing blocks or non-linearities in the processing chain, or failure of other models (e.g., background noise estimates). This problem becomes more challenging during full-duplex modes of communication where echoes from double talk scenarios are difficult to suppress without significant distortion or attenuation \cite{noauthor_ieee_2010}. 

With the advent of deep learning techniques, many supervised learning algorithms for AEC have shown better performance compared to their classical counterparts, e.g., \cite{fazel_cad-aec_2020, halimeh_efficient_2020, ma_acoustic_2020}. Some studies have also shown good performance using a combination of classical and deep learning methods such as using adaptive filters and \emph{recurrent neural networks} (RNNs) \cite{ma_acoustic_2020, zhang_deep_2019} but only on synthetic datasets. While these approaches are promising, they lack evidence of their performance on real-world datasets with speech recorded in diverse noise and reverberant environments. This makes it difficult for researchers in the industry to choose a good model that can perform well on a representative real-world dataset.

Most AEC publications use objective measures such as \emph{echo return loss enhancement} (ERLE) \cite{enzner_acoustic_2014} and \emph{perceptual evaluation of speech quality} (PESQ) \cite{noauthor_itu-t_2001}. ERLE in dB is defined as: 

\begin{equation}
ERLE = 10\log_{10} \frac{\mathbb{E}[y^2(n)]}{\mathbb{E}[e^2(n)]} 
\end{equation}

\noindent where $y(n)$ is the microphone signal, and $e(n)$ is the residual echo after cancellation. ERLE is only appropriate when measured in a quiet room with no background noise and only for single talk scenarios (not double talk), where we can use the processed microphone signal as an estimate for $e(n)$. PESQ has also been shown to not have a high correlation to subjective speech quality in the presence of background noise \cite{avila_non-intrusive_2019}. Using the datasets provided in this challenge we show that ERLE and PESQ have a low correlation to subjective tests (Table \ref{tab:correlation}). In order to use a dataset with recordings in real environments, we can not use ERLE and PESQ. A more reliable and robust evaluation framework is needed that everyone in the research community can use, which we provide as part of the challenge.

\begin{table}
\centering
\caption{Pearson Correlation Coefficient (PCC) and Spearman's Rank Correlation Coefficient (SRCC) between objective and subjective P.808 results on single talk echo scenarios (see Section \ref{sec:framework}).}
\label{tab:correlation}
\begin{tabular}{ccc}
\toprule
{} & PCC & SRCC \\
\midrule
ERLE & 0.31 & 0.23 \\
PESQ & 0.67 & 0.57 \\
\bottomrule
\end{tabular}
\end{table}

This AEC challenge is designed to stimulate research in the AEC domain by open sourcing a large training dataset, test set, and subjective evaluation framework. We provide two new open source datasets for training AEC models. The first is a real dataset captured using a large-scale crowdsourcing effort. This dataset consists of real recordings that have been collected from over 10,000 diverse audio devices and environments. The second dataset is synthesized from speech recordings, room impulse responses, and background noise derived from \cite{reddy_interspeech_2020}. An initial test set will be released for the researchers to use during development and a blind test set near the end, which will be used to decide the final competition winners. We believe these datasets are large enough to facilitate deep learning and representative enough for practical usage in shipping telecommunication products.

This is the third AEC challenge we have conducted. The first challenge was held at ICASSP 2021 \cite{sridhar_icassp_2021} and the second at INTERSPEECH 2021 \cite{cutler_interspeech_2021}. These challenges had 31 participants with entries ranging from pure deep models, hybrid linear AEC + deep echo suppression, and DSP methods. The results show that the deep and hybrid models far outperformed DSP methods, with the latest winners being both pure deep and hybrid models. However, there is still much room for improvement. To improve the challenge and further stimulate research in this area we have made the following changes:
\begin{itemize}
    \item The dataset has increased from 5,000 devices and environments to 10,000 to provide additional training data.
    \item Mobile phone scenarios are now included, which are an important area that is even more challenging than desktop or notebook computers. 50\% of the blind test set were mobile devices, and 50\% were desktop devices. 
    \item The Microsoft Speech Recognizer's Word Accuracy rate (WAcc) is used as a metric in the challenge, as many scenarios include speech recognition and the AEC should not degrade WAcc. WAcc = 1 - Word Error Rate.
    \item The test sets are now 48 kHz, which is an important requirement for many scenarios.
\end{itemize}

The training dataset is described in Section \ref{sec:data}, and the test set in Section \ref{ssec:data_test}. We describe a DNN-based AEC method in Section \ref{sec:model}. The online subjective evaluation framework is discussed in Section \ref{sec:framework}, and the objective service in Section \ref{sec:metric}. The challenge metric is given in Section \ref{sec:challenge_metric} and the challenge rules are described in \url{https://aka.ms/aec-challenge}.

\section{Training datasets}
\label{sec:data}
The challenge will include two new open source datasets, one real and one synthetic. The datasets are available at \url{https://github.com/microsoft/AEC-Challenge}.

\subsection{Real dataset}
\label{ssec:real_data}
 The first dataset was captured using a large-scale crowdsourcing effort. This dataset consists of more than 50,000 recordings from over 10,000 different real environments, audio devices, and human speakers in the following scenarios:

\begin{enumerate}
    \item Far end single talk, no echo path change
    \item Far end single talk, echo path change
    \item Near end single talk, no echo path change
    \item Double talk, no echo path change
    \item Double talk, echo path change
    \item Sweep signal for RT60 estimation
\end{enumerate}

\begin{figure}[t]
    \centering
    \includegraphics[width=160pt]{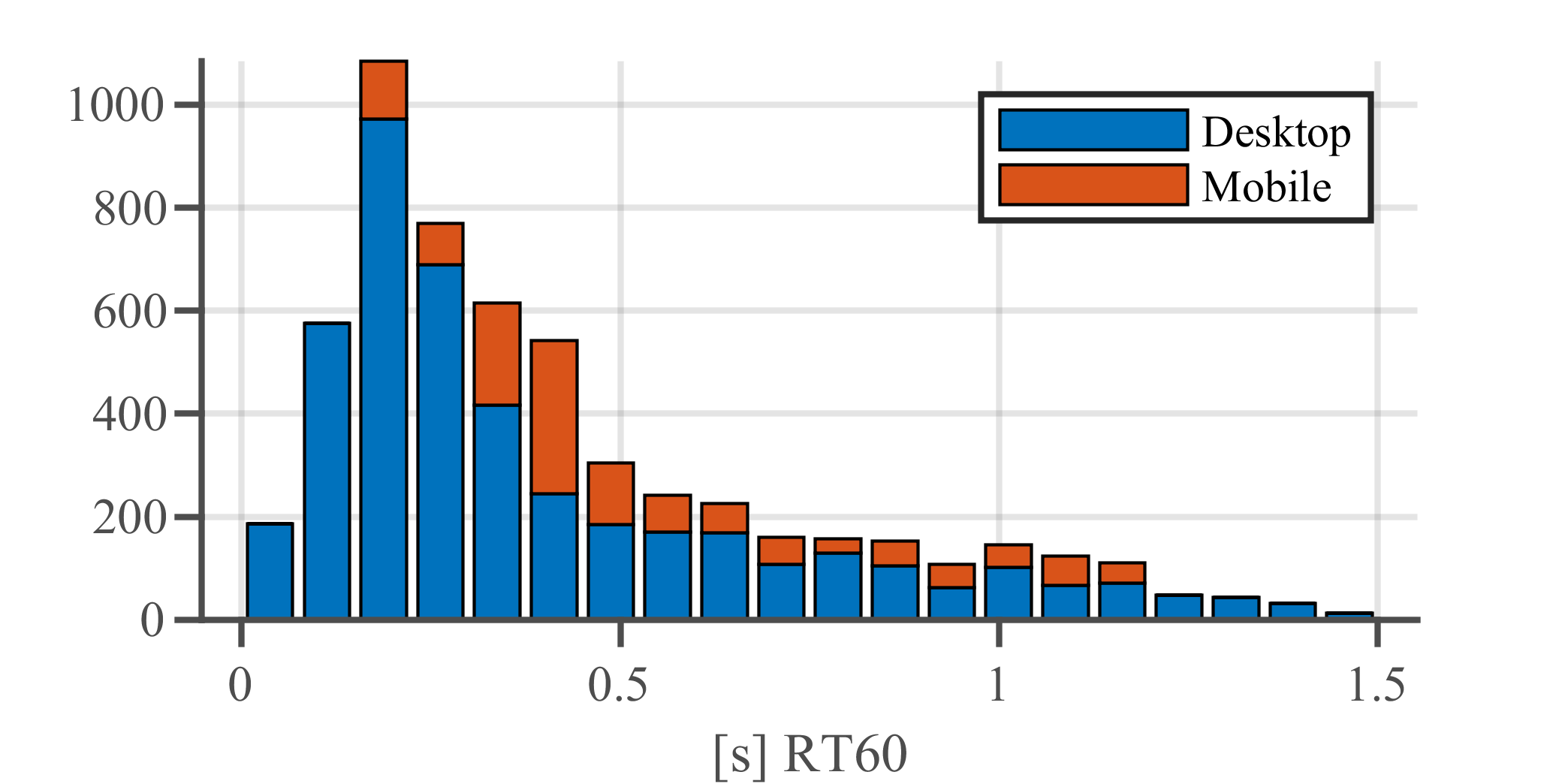}
    \caption{Distribution of reverberation time (RT60).}
    \label{fig:rt60}
\end{figure}

For the far end single talk case, there is only the loudspeaker signal (far end) played back to the users and users remain silent (no near end speech). For the near end single talk case, there is no far end signal and users are prompted to speak, capturing the near end signal. For double talk, both the far end and near end signals are active, where a loudspeaker signal is played and users talk at the same time. Echo path changes were incorporated by instructing the users to move their device around or bring themselves to move around the device.  The RT60 distribution for 4387 desktop environments in the real dataset for which impulse response measurements were available is estimated using a method by Karjalainen et al.~\cite{karjalainen_estimation_2001} and shown in Figure \ref{fig:rt60}. For 1251 mobile environments the RT60 distribution shown was estimated blindly from speech recordings~\cite{gamper_blind_2018}.%The RT60 estimates can be used to sample the dataset for training.

% The near end single talk speech quality is given in Figure \ref{fig:nearend}.

\begin{figure}[t]
    \centering
    \includegraphics[width=200pt]{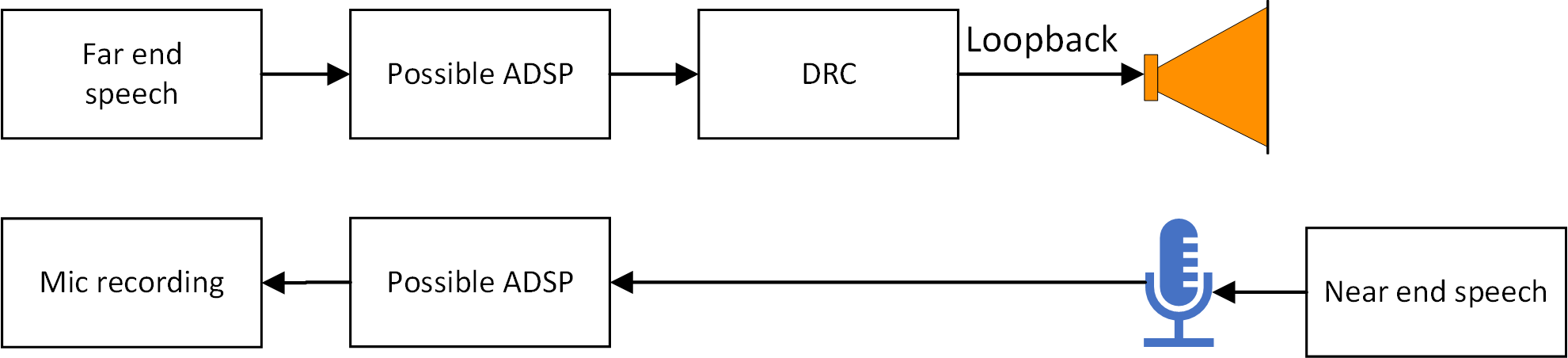}
    \caption{The custom recording application recorded the loopback and microphone signals.}
    \label{fig:recording}
\end{figure}

We use \emph{Amazon Mechanical Turk} as the crowdsourcing platform and wrote a custom HIT application that includes a custom tool that users download and execute to record the six scenarios described above. The dataset includes Microsoft Windows and Android devices. Each scenario includes the microphone and loopback signal (see Figure \ref{fig:recording}). Even though our application uses the WASAPI raw audio mode to bypass built-in audio effects, the PC can still include Audio DSP on the receive signal (e.g., equalization and Dynamic Range Compression (DRC)); it can also include Audio DSP on the send signal, such as AEC and noise suppression.

For far end signals, we use both clean speech and real world recordings. For clean speech far end signals, we use the speech segments from the Edinburgh dataset \cite{valentini-botinhao_speech_2016}. This corpus consists of short single speaker speech segments ($1$ to $3$ seconds). We used a \emph{long short term memory} (LSTM) based gender detector to select an equal number of male and female speaker segments. Further, we combined $3$ to $5$ of these short segments to create clips of length between $9$ and $15$ seconds in duration. Each clip consists of a single gender speaker. We create a gender-balanced far end signal source comprising of $500$ male and $500$ female clips. Recordings are saved at the maximum sampling rate supported by the device and in 32-bit floating point format; in the released dataset we down-sample to 48 kHz and 16-bit using automatic gain control to minimize clipping.

For noisy speech far end signals we use $2000$ clips from the near end single talk scenario. Clips are gender balanced to include an equal number of male and female voices.

For the far end single talk scenario, the clip is played back twice. This way, the echo canceller can be evaluated both on the first segment, when it has had minimal time to converge, and on the second segment, when the echo canceller has converged and the result is more indicative to a real call scenario.

For the double talk scenario, the far end signal is similarly played back twice, but with an additional silent segment in the middle, when only near end single talk occurs.

For near end speech, the users were prompted to read sentences from a TIMIT \cite{garofolo_darpa_1993} sentence list. Approximately 10 seconds of audio is recorded while the users are reading.

\if 0
\begin{figure}[t]
    \centering
    \includegraphics[width=160pt]{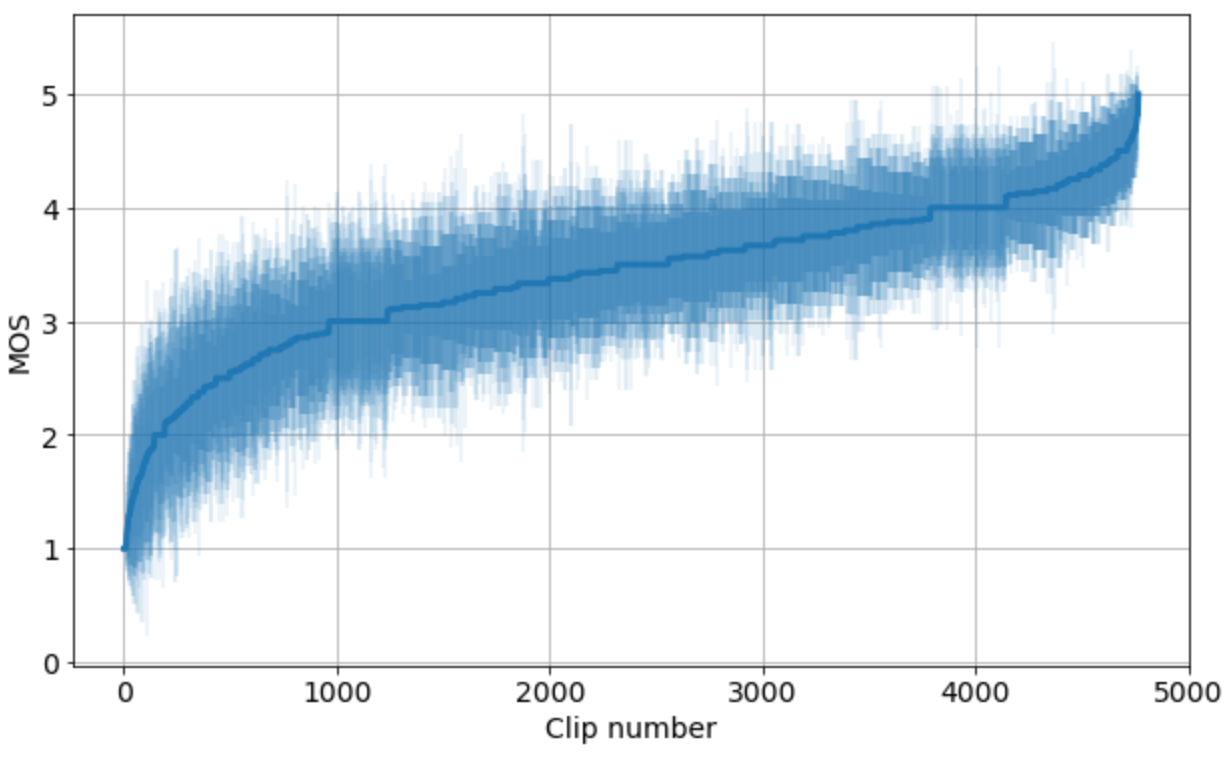}
    \caption{Sorted near end single talk clip quality (P.808) with 95\% confidence intervals.}
    \label{fig:nearend}
\end{figure}
\fi

\subsection{Synthetic dataset}
\label{ssec:synth_data}
The second dataset provides 10,000 synthetic scenarios, each including single talk, double talk, near end noise, far end noise, and various nonlinear distortion scenarios.
Each scenario includes a far end speech, echo signal, near end speech, and near end microphone signal clip.
We use 12,000 cases (100 hours of audio) from both the clean and noisy speech datasets derived in \cite{reddy_interspeech_2020} from the LibriVox project\footnote{\url{https://librivox.org}} as source clips to sample far end and near end signals.
The LibriVox project is a collection of public domain audiobooks read by volunteers.
\cite{reddy_interspeech_2020} used the online subjective test framework ITU-T P.808 to select audio recordings of good quality (4.3 $\leq$ MOS $\leq$ 5) from the LibriVox project.
The noisy speech dataset was created by mixing clean speech with noise clips sampled from Audioset \cite{gemmeke_audio_2017}, Freesound\footnote{\url{https://freesound.org}} and DEMAND \cite{thiemann_diverse_2013} databases at signal to noise ratios sampled uniformly from [0, 40] dB.

To simulate a far end signal, we pick a random speaker from a pool of 1,627 speakers, randomly choose one of the clips from the speaker, and sample 10 seconds of audio from the clip.
For the near end signal, we randomly choose another speaker and take 3-7 seconds of audio which is then zero-padded to 10 seconds.
Of the selected far end and near end speakers, 71\% and 67\% are male, respectively.
To generate an echo, we convolve a randomly chosen room impulse response from a large internal database with the far end signal. The room impulse responses are generated by using Project Acoustics technology\footnote{\url{https://www.aka.ms/acoustics}} and the RT60 ranges from 200 ms to 1200 ms.
In 80\% of the cases, the far end signal is processed by a nonlinear function to mimic loudspeaker distortion. 
For example, the transformation can be clipping the maximum amplitude, using a sigmoidal function as in \cite{lee_dnn-based_2015}, or applying learned distortion functions, the details of which we will describe in a future paper.
This signal gets mixed with the near end signal at a signal to echo ratio uniformly sampled from -10 dB to 10 dB. The signal to echo ratio is calculated based on the clean speech signal (i.e., a signal without near end noise).
The far end and near end signals are taken from the noisy dataset in 50\% of the cases.
The first 500 clips can be used for validation as these have a separate list of speakers and room impulse responses.
Detailed metadata information can be found in the repository.

% todo: percent of data with cascaded AEC, headsets

\section{Test set}
\label{ssec:data_test}

Two test sets are included, one at the beginning of the challenge and a blind test set near the end. Both consist of 800 real world recordings, between ~30-45 seconds in duration. The datasets include the following scenarios that make echo cancellation more challenging:

\begin{itemize}
    \item Long- or varying delays, i.e., files where the delay between loopback and mic-in is atypically long or varies during the recording.
    \item Strong speaker and/or mic distortions.
    \item Stationary near end noise
    \item Non-stationary near end noise
    \item Recordings with audio DSP processing from the device, such as AEC or noise reduction
    \item Glitches, i.e., files with ``choppy" audio, for example, due to very high CPU usage
    \item Gain variations, i.e., recordings where far end level changes during the recording
  (\ref{ssec:real_data}), sampled randomly
\end{itemize}

% We use the method described in \cite{gitiaux_aura_2021} to make the test set more challenging and representative of the dataset. 

\section{Baseline AEC Method}
\label{sec:model}
We adapt a noise suppression model developed in \cite{xia_weighted_2020} to the task of echo cancellation. 
Specifically, a recurrent neural network with gated recurrent units takes concatenated log power spectral features of the microphone signal and far end signal as input, and outputs a spectral suppression mask. 
The short-time Fourier transform is computed based on 20 ms frames with a hop size of 10 ms, and a 320-point discrete Fourier transform.
We use a stack of two gated recurrent unit layers, each of size 322 nodes, followed by a fully-connected layer with a sigmoid activation function. The model has 1.3 million parameters.
The estimated mask is point-wise multiplied with the magnitude spectrogram of the microphone signal to suppress the far end signal. 
Finally, to resynthesize the enhanced signal, an inverse short-time Fourier transform is used on the phase of the microphone signal and the estimated magnitude spectrogram.
We use a mean squared error loss between the clean and enhanced magnitude spectrograms. 
The Adam optimizer with a learning rate of 0.0003 is used to train the model. The model and the inference code is available in the challenge repository.\footnote{\url{https://github.com/microsoft/AEC-Challenge/tree/main/baseline/icassp2022}}

\section{Online subjective evaluation framework}
\label{sec:framework}
We have extended the open source P.808 Toolkit \cite{naderi_open_2020} with methods for evaluating the echo impairments in subjective tests. We followed the \textit{Third-party Listening Test B} from ITU-T Rec. P.831 \cite{noauthor_itu-t_1998-1} and ITU-T Rec. P.832 \cite{noauthor_itu-t_2000} and adapted them to our use case as well as for the crowdsourcing approach based on the ITU-T Rec. P.808 \cite{noauthor_itu-t_2018} guidance. 

A third-party listening test differs from the typical listening-only tests (according to the ITU-T Rec. P.831) in the way that listeners hear the recordings from the \textit{center} of the connection rather in the former one in which the listener is positioned at one end of the connection \cite{noauthor_itu-t_1998-1}. Thus, the speech material should be recorded by having this concept in mind.
During the test session, we use different combinations of single- and multi-scale Absolute Category Ratings depending on the speech sample under evaluation. We distinguish between single talk and double talk scenarios.
For the near end single talk, we ask for the overall quality. For the far end single talk and double talk scenario, we ask for an echo annoyance and for impairments of other degradations in two separate questions\footnote{Question 1: How would you judge the degradation from the echo? Question 2: How would you judge other degradations (noise, missing audio, distortions, cut-outs)?}. Both impairments are rated on the degradation category scale (from 1:\textit{Very annoying}, to 5:\textit{Imperceptible}) to obtain Degradation Mean Opinion Scores (DMOS). Note that we do not use the Other degradation category for far end single talk for evaluating echo cancellation performance, since this metric mostly reflects the quality of the original far end signal. However, we have found that having this component in the questionnaire helps increase the accuracy of echo degradation ratings (when measured against expert raters). Without the Other category, raters can sometimes assign degradations due to noise to the Echo category.

For the far end single talk scenario, we evaluate the second half of each clip to avoid initial degradations from initialization, convergence periods, and initial delay estimation. For the double talk scenario, we evaluate roughly the final third of the audio clip.

The subjective test framework with an AEC extension is available at \url{https://github.com/microsoft/P.808}. A more detailed description of the test framework and its validation is given in \cite{cutler_crowdsourcing_2020}.

\section{Azure service objective metric}
\label{sec:metric}
We have developed an objective perceptual speech quality metric called AECMOS. It can be used to stack rank different AEC methods based on MOS estimates with high accuracy. It is a neural network-based model that is trained using the ground truth human ratings obtained using our online subjective evaluation framework. The audio data used to train the AECMOS model is gathered from the numerous subjective tests that we conducted in the process of improving the quality of our AECs as well as the first two AEC challenge results. The performance of AECMOS on AEC models is given in Table~\ref{tab:AECMOS} compared with subjective human ratings on the 18 submitted models. We note that this model had not seen any mobile nor fullband data during training. The next version of AECMOS will have mobile and fullband data in its training data. A more detailed description of AECMOS is given in \cite{purin_aecmos_2022}. Sample code and details of the evaluation API can be found on \url{https://aka.ms/aec-challenge}.

\begin{table}[tb]
\centering
\caption{AECMOS PCC and SRCC} %model 1251 results
\label{tab:AECMOS}
\begin{tabular}{c c c c c} 
\toprule
Scenario & PCC  & SRCC \\
\midrule
Far end single talk echo DMOS & 0.828  & 0.719 \\ 
Near end single talk MOS & 0.843 & 0.856\\
Double talk echo DMOS & 0.882 & 0.766 \\ 
Double talk other DMOS & 0.929 & 0.913\\ 
\bottomrule
\end{tabular}
\end{table}

\section{Challenge metric}
\label{sec:challenge_metric}
The challenge performance is determined using the average of the four subjective scores described in Section \ref{sec:framework} and WAcc, all weighted equally. Specifically:

\noindent $M = \frac{\frac{(FE_{ST}-1)}{4} + \frac{(NE_{ST} - 1)}{4} + \frac{(DT_{echo}-1)}{4}  + \frac{(DT_{other}-1)}{4} + WAcc}{5}$

\noindent where $FE_{ST}$ is far end single talk, $NE_{ST}$ is near end single talk, $DT_{echo}$ is double talk echo, and $DT_{other}$ is double talk other.

\section{Results}
\label{sec:results}
We received 18 submissions for the challenge. Each team submitted processed files from the blind test set (see Section \ref{ssec:data_test}). 
We batched all submissions into three sets:
 \begin{itemize}
  \item Near end single talk files for a MOS test (NE ST MOS).
  \item Far end single talk files for an Echo and Other degradation DMOS test (FE ST Echo/Other DMOS).
  \item Double talk files for an Echo and Other degradation DMOS test (DT Echo/Other DMOS).
\end{itemize}

\begin{figure}[t]
    \centering
    \includegraphics[width=240pt]{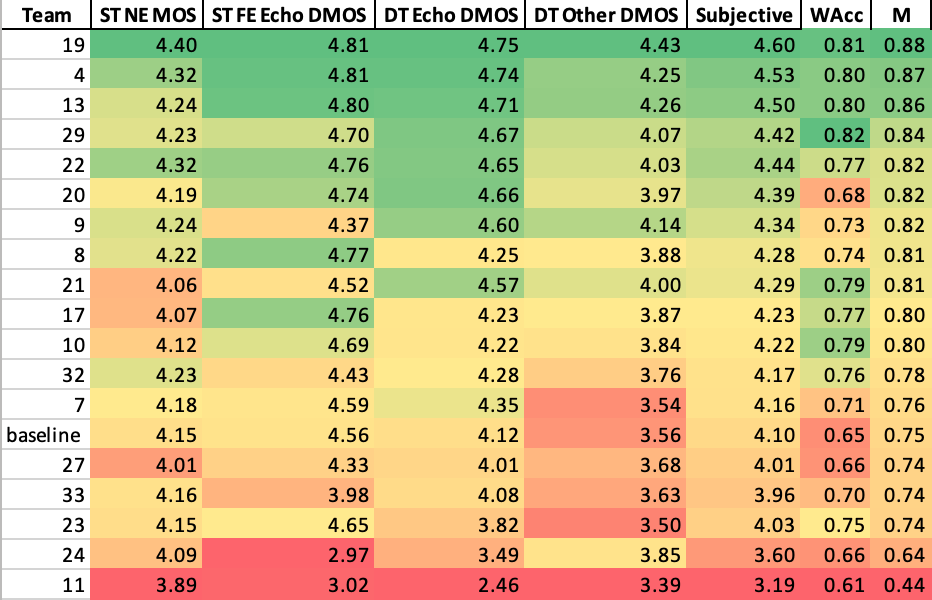}
    \caption{AEC challenge results}
    \label{fig:results}
\end{figure}

The results are shown in Figure \ref{fig:results}. The score differences in near end, echo, double talk, and WAcc  highlight the importance of evaluating all scenarios, since in many cases, performance in one scenario comes at a cost in another scenario. The PCC of WAcc and the mean subjective scores is 0.85, which helps motivate why WAcc needs to be measured.

For the top performing teams, we ran an ANOVA test to determine statistical significance (see \url{https://aka.ms/aec-challenge}). The 2nd and 3rd, and 5th and 6th places were tied. For the ties, the winners were selected using the lower complexity model.  

A high-level comparison of the top 5 performing models is given in Table~\ref{tab:Top5}. Real-time factor is the run-time / frame time on an Intel Core i5 quad core 2.4 GHz CPU or equivalent. The 1st place model \cite{zhang_multi-scale_2022} also won the ICASSP 2022 DNS Challenge \cite{dubey_icassp_2022}, providing the only model that didn't induce SIG \cite{naderi_subjective_2021} distortion in that challenge. It is a hybrid model but is unique in that it uses the linear AEC only to condition the DNN, not filtering the audio. Three of the top 5 teams use linear AEC's and DNN's, and all 5 use the STFT domain. In addition, all 5 models perform noise suppression in addition to AEC. There is a wide range of model sizes and complexities, and it wasn't necessary to use external datasets to do well in the challenge. 

\begin{figure}[t]
    \centering
    \includegraphics[width=240pt]{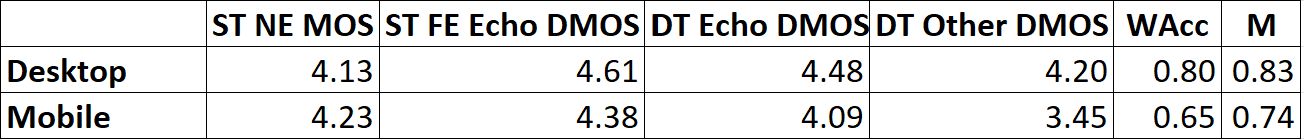}
    \caption{Results for desktop and mobile recordings}
    \label{fig:desktop_vs_mobile}
\end{figure}

\begin{table}[tb]
\centering
\caption{Comparison of the top 5 teams} 
\label{tab:Top5}
\begin{tabular}{c c c r c c} 
\toprule
Place & Paper & Hybrid & Params & Real-time & Additional \\
  & &  &  & Factor & Datasets   \\\midrule
1 & \cite{zhang_multi-scale_2022} & Y* & 1.5 M & 0.60 & N \\ 
2 & \cite{zhao_deep_2022} & Y & 17.4 M & 0.10 & Y \\ 
3 & \cite{zhang_multi-task_2022} & Y & 4.8 M & 0.20 & Y \\
4 & \cite{sun_explore_2022} & Y & 55.5 M & 0.30 & N \\
5 & \cite{cui_multi-scale_2022} & N & 4.3 M & 0.02 & Y \\
\bottomrule
\end{tabular}
\end{table}

When comparing the results between mobile and desktop recordings (Figure \ref{fig:desktop_vs_mobile}), we observe relatively similar scores for the near end single talk category, but significantly lower scores for mobile in echo categories, especially for double talk. The difference is highest in the double talk degradation category, where the score for mobile recordings is lower by 0.75 MOS. One reason for this is that in mobile devices, the loudspeaker is closer to the microphone, so the signal-to-echo ratio in these devices is lower on average.

\section{Conclusions}
\label{sec:end}
While the results of this challenge continue to improve over previous challenges, there is still significant room for improvement, especially with the mobile scenario.  We hope this challenge, dataset, test set, and test framework stimulate research in this important area of speech enhancement. 

% To start a new column (but not a new page) and help balance the last-page
% column length use \vfill\pagebreak.
% -------------------------------------------------------------------------
%\vfill\pagebreak
\pagebreak

% References should be produced using the bibtex program from suitable
% BiBTeX files (here: strings, refs, manuals). The IEEEbib.bst bibliography
% style file from IEEE produces unsorted bibliography list.
% -------------------------------------------------------------------------
\bibliographystyle{IEEEbib}
\bibliography{IC3-AI}
%\bibliography{references}
\end{document}